\documentclass[smallextended,natbib]{svjour3}
\usepackage[switch]{lineno}

\usepackage{graphicx}
\usepackage{amsmath,amssymb,mathtools}
\usepackage{bm}
\usepackage{booktabs}
\usepackage{siunitx}

\usepackage{tikz}
\usetikzlibrary{arrows.meta,positioning}

\usepackage{xcolor}
\usepackage[hidelinks]{hyperref}

\journalname{Earth Science Informatics}

\newcommand{\R}{\mathbb{R}}

\title{Koopman Analysis of Sea Surface Temperature with a Signature Kernel}
\titlerunning{Signature-kernel Koopman Analysis}

\author{Nozomi Sugiura \and Satoshi Osafune \and Shinya Kouketsu}

\authorrunning{N. Sugiura et al.} 

\institute{
N. Sugiura \at
Japan Agency for Marine-Earth Science and Technology, Yokosuka, Japan \\
\email{nsugiura@jamstec.go.jp}
\and
Satoshi Osafune \at
Japan Agency for Marine-Earth Science and Technology, Yokosuka, Japan 
\and 
Shinya Kouketsu \at
Japan Agency for Marine-Earth Science and Technology, Yokosuka, Japan \\
Advanced Institute for Marine Ecosystem Change (WPI-AIMEC), Yokohama, Japan.
}

\graphicspath{{figs/}}

\begin{document}
\maketitle

\begin{abstract}
We develop a trajectory-based Koopman method for sea surface temperature (SST) that lifts annual SST segments using a
signature kernel---a reproducing kernel Hilbert space (RKHS) kernel that compares paths via iterated-integral features---and learns the one-year shift operator.
By operating on annual trajectory segments rather than instantaneous fields, the method encodes finite-time history, which helps capture memory effects in SST-only evolution. The resulting operator improves out-of-sample multi-year forecast skill
relative to a climatology baseline and reveals coherent spectral modes. We implement the approach via kernel extended dynamic mode decomposition (EDMD) on
signature-kernel Gram matrices, yielding a single pipeline for forecasting and spectral diagnostics of high-dimensional SST
dynamics.
\end{abstract}

\section{Introduction}\label{sec:intro}
Sea surface temperature (SST) is a primary manifestation of air--sea interaction and a central observable parameter
for characterizing climate variability across a wide range of time scales \citep[e.g.,][]{deser2010sea}.
A common starting point in statistical SST analysis and prediction is to treat the evolution of snapshot SST fields (often monthly means)
as approximately Markovian and amenable to linear or linearized modeling, including linear decompositions and related statistical learning approaches \citep[e.g.,][]{Penland1993,newman2013empirical,Ho2013,Bilgili2024}.
However, when only a subset of the underlying ocean--atmosphere state is observed (e.g., SST alone),
the induced dynamics of the observed variables is generally history-dependent (effectively non-Markovian) and can be nonlinear due to the intrinsic nature of the underlying dynamics.
These two departures---memory and nonlinearity---motivate 
enlarging the state along the time axis
and seeking a linear description of evolution by adopting a functional viewpoint, rather than relying on linearization.

To account for the memory effect, we work with trajectories rather than snapshots, using a time-ordered sequence of annual path segments as the state.
In contrast to Takens-style delay-coordinate embeddings, which represent history as a delay vector \citep[e.g.,][]{Takens1981},
we retain an explicit order structure and treat the state as an ordered path (trajectory) object.

To account for nonlinearity while keeping a linear evolution law, we describe temporal evolution by a linear operator acting on
\emph{observables}, i.e., scalar-valued functions of a path.
Crucially, by lifting the dynamics to a sufficiently rich space of observables, nonlinear state evolution can be represented (or well-approximated) by a linear operator.
This perspective leads naturally to Koopman analysis \citep[e.g.,][]{mezic2013analysis,williams2015data} on path space.

A remaining practical question is how to construct a rich yet tractable function family on path space.
Path signatures provide a systematic lifting that encodes temporal order and generates expressive features from trajectories \citep[e.g.,][]{lyons2014rough}.
Recent work has pursued tractable Koopman learning using truncated signature features \citep[e.g.,][]{Chretien2026GSI}; such explicit truncations are most practical for low-dimensional paths and moderate truncation levels, whereas kernelization of signatures \citep[e.g.,][]{Kiraly2019} offers an RKHS route that scales to high-dimensional spatiotemporal fields \citep[e.g.,][]{williams2015}.
Motivated by high-dimensional climate fields such as sea surface temperature (SST), we adopt a signature-\emph{kernel} kEDMD formulation for Koopman-operator learning.

In this study, we propose a procedure to learn an explicit linear Koopman operator for the one-step shift of annual SST paths via signature-kernel EDMD (Fig.~\ref{fig:pipeline}),
yielding a single operator representation that supports both out-of-sample forecasting and spectral (mode) diagnostics.

\subsection*{Contributions}
\begin{itemize}
\item \textbf{Trajectory state representation for SST-only dynamics.}
To account for effective non-Markovianity under partial observation, we represent the record as annual trajectory segments (paths) and formulate Koopman analysis on this path space.

\item \textbf{Signature-kernel lifting for tractable Koopman learning on paths.}
To obtain a linear-operator description in the presence of nonlinear dynamics, we learn an explicit Koopman operator for the one-year path shift using signature-kernel EDMD, thereby providing a systematic and scalable function family of \emph{observables} on path space.

\item \textbf{Strict time-ordered evaluation and diagnostics from a single estimator.}
Using only information available up to each forecast time, we demonstrate out-of-sample multi-year forecast skill relative to climatology and a naive kernel-EDMD baseline, with all preprocessing and model selection performed without access to future data.
The same learned operator also yields coherent spectral diagnostics (eigenvalues, eigenfunctions, and Koopman modes).
\end{itemize}

\section{Background and related work}\label{sec:related}

\paragraph{Koopman operators and data-driven approximations.}
The Koopman operator provides a linear operator-theoretic description of nonlinear dynamics by acting on
observables rather than states \citep{Koopman1931}. Data-driven realizations of this concept are now widely
used in data science and fluid dynamics, notably through dynamic mode decomposition and its extensions
\citep{mezic2013analysis}. EDMD and its kernelized variant (kEDMD)
offer systematic finite-dimensional approximations of the Koopman operator by projecting onto a chosen
function space or a reproducing kernel Hilbert space (RKHS) \citep{williams2015data,williams2015}.

\paragraph{Ocean and SST dynamics: from linear Markov models to Koopman.}
A classical approach to extracting dynamical structure from SST is the linear inverse model (LIM) developed by
Penland and collaborators \citep{Penland1993,Penland1995}, which models SST anomalies as a linear Markov
process. While LIM is influential, it assumes linearity and Markovianity in the observed SST variables,
even though SST alone does not constitute a closed state of the coupled ocean--atmosphere system.
More recently, Koopman-based analyses of SST using kEDMD estimate an explicit linear operator for month-to-month evolution
and extract coherent spectral modes at interannual scales \citep{Navarra2021,Navarra2024,LorenzoSanches2025}.
These studies motivate operator-based diagnostics while highlighting the need for formulations that more effectively encode
memory effects in SST-only evolution.

\paragraph{Neural operators as predictive baselines and their limitations for mode diagnostics.}
Recent progress in data-driven forecasting has been driven by neural operators (e.g., Fourier neural operators), which learn mappings between input and output fields and can often transfer across grid resolutions.
Related progress has also been made with deep learning and time-series models for monthly SST prediction \citep[e.g.,][]{Jia2024}.
For example, OceanNet \citep{Chattopadhyay2024} demonstrates competitive seasonal prediction skill for regional ocean dynamics.
Although neural-operator layers may involve linear integral (or Fourier) operators, the overall predictor is typically nonlinear due to compositions with nonlinear components, so Koopman-type linear spectral diagnostics (eigenvalues/eigenfunctions/modes) are not usually obtained as a native output and often require additional post-processing.
In contrast, Koopman/kEDMD estimates an explicit linear operator, from which spectral modes follow directly alongside prediction.

\paragraph{Why pathwise representations are needed.}
In SST-only settings, incorporating temporal context is essential for learning effective dynamics.
Time-delay coordinates are a standard remedy \citep{Takens1981}. By contrast, path signatures provide a canonical,
order-sensitive collection of iterated-integral features for trajectories \citep{lyons1998differential,lyons2014rough}.
Signature kernels provide an RKHS route that retains pathwise information while enabling Gram-matrix computations
\citep{Kiraly2019}.

\paragraph{Gap addressed by this work.}
We learn the Koopman operator of the one-year path shift on annual SST trajectory segments using signature-kernel kEDMD, obtaining an explicit operator that supports both strictly out-of-sample forecasting and spectral-mode diagnostics.

\section{Methods}\label{sec:methods}

We represent each year as a path $X_t$ and model the dynamics as the one-year path shift $F(X_t)=X_{t+1}$.
We consider observables $g$ on the annual-path set and approximate the Koopman action
$(\mathcal{K}g)(X)=g(F(X))$ in an RKHS induced by a path kernel.
Kernel EDMD yields a finite-dimensional matrix $K$ whose spectrum and iterates approximate those of $\mathcal{K}$,
enabling both spectral diagnostics and multi-year forecasting.

\subsection{Pipeline overview}\label{sec:pipeline}
\begin{figure}[t]
\centering
\begin{tikzpicture}[
  font=\small,
  box/.style={
    draw, rounded corners=2pt, align=center,
    inner sep=4pt, minimum height=9mm,
    text width=0.92\linewidth
  },
  sbox/.style={
    draw, rounded corners=2pt, align=center,
    inner sep=4pt, minimum height=9mm,
    text width=0.44\linewidth
  },
  arrow/.style={-Latex, thick},
  node distance=5mm
]

\node[box] (prep) {Monthly SST $\{x_m\}$\\
$\rightarrow$ past-only rolling climatology $\rightarrow$ anomalies $\{x'_m\}$\\
$\rightarrow$ annual paths $\{X_t\}$ (cumulative sums of $\{x'_m\}$)};

\node[box, below=of prep] (ksig) {Path kernel\\
main: $\kappa_{\mathrm{sig}}^{(n,\lambda)}$ (signature kernel)\\
baseline: $\kappa_{\mathrm{SPK}}$ (month-aligned baseline)};

\node[box, below=of ksig] (GA) {Gram and cross-Gram matrices\\
$G$ and $A$};

\node[box, below=of GA] (kedmd) {kEDMD on annual paths\\
$\Rightarrow$ estimate a Koopman matrix $K$};

\node[sbox, below left=4mm and 0mm of kedmd.south] (lfo)
{LFO: forecast-skill evaluation};

\node[sbox, below right=4mm and 0mm of kedmd.south] (lso)
{LSO: spectral diagnostics};

\draw[arrow] (prep) -- (ksig);
\draw[arrow] (ksig) -- (GA);
\draw[arrow] (GA) -- (kedmd);
\draw[arrow] (kedmd) -- (lfo.north);
\draw[arrow] (kedmd) -- (lso.north);

\end{tikzpicture}
\caption{Overview of the trajectory-based Koopman pipeline via signature-kernel kEDMD.}
\label{fig:pipeline}
\end{figure}
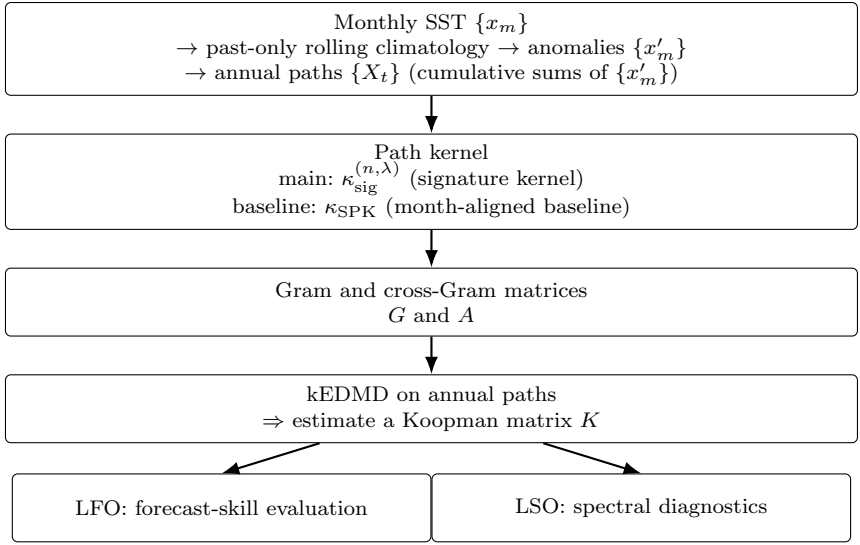

Figure~\ref{fig:pipeline} summarizes the workflow.
We preprocess monthly SST fields into anomalies using a strictly past-only rolling climatology and partition the record
into annual path segments (Section~\ref{sec:preprocess_path} and Appendix~\ref{app:preprocess}).
Annual segments are compared by a path kernel: our main choice is the truncated signature kernel
$\kappa_{\mathrm{sig}}^{(n,\lambda)}$, and we also use a snapshot-aligned Sum-of-Pairs Kernel (SPK) baseline
(Section~\ref{sec:koopman_full}). From the resulting Gram and cross-Gram matrices, we apply kEDMD to estimate a
finite-dimensional Koopman matrix $K$.
Evaluation protocols are defined in Section~\ref{sec:protocols}; experimental settings are summarized in
Section~\ref{sec:exp}.

\subsection{Annual path segments from monthly anomalies}\label{sec:preprocess_path}

Monthly SST fields $\{x_m\}$ are converted into anomalies $\{x'_m\}$ using a strictly past-only rolling monthly climatology
(Appendix~\ref{app:preprocess}).
Fix a start month $m_0$ and group 12 consecutive anomalies $\{x'_{m_0+12t+i}\}_{i=0}^{11}$ to form an annual segment for each year index $t$.
Each segment is represented as a piecewise-linear path $X_t:[0,1]\to\R^{d}$ obtained by cumulative summation of within-year anomalies.

Let $\mathcal{P}:=\{X_t\}_{t\in\mathcal{T}}\subset C([0,1],\R^{d})$ be the set of available annual paths.
The one-year shift is defined on successor pairs by $F(X_t):=X_{t+1}$, and the sampled Koopman action is
\[
(\mathcal{K}g)(X):=g(F(X)).
\]

\subsection{Signature-kernel lifting on path space}\label{sec:sigkernel}

Let $k:\R^{d}\times\R^{d}\to\R$ be a positive semidefinite \emph{base kernel} with RKHS $\mathcal{H}_k$ and feature map
$\phi(x):=k(x,\cdot)\in\mathcal{H}_k$. Lift a segment $X_t$ to an $\mathcal{H}_k$-valued path
$\Phi_t(u):=\phi(X_t(u))$.
For $\ell\ge 1$, define the level-$\ell$ signature tensor of $\Phi_t$ by
\begin{equation}\label{eq:signature_level}
S^{(\ell)}(\Phi_t)
:=\int_{0<u_1<\cdots<u_\ell<1} d\Phi_t(u_1)\otimes\cdots\otimes d\Phi_t(u_\ell)
\in \mathcal{H}_k^{\otimes \ell},
\end{equation}
and set $S^{(0)}(\Phi_t):=1$.

\paragraph{Signature kernel.}
For truncation level $n\in\mathbb{N}$ and dilation $\lambda>0$, define
\begin{equation}\label{eq:sig_kernel}
\kappa_{\mathrm{sig}}^{(n,\lambda)}(X_s,X_t)
:=\sum_{\ell=0}^{n}\lambda^{2\ell}\,
\bigl\langle S^{(\ell)}(\Phi_s),S^{(\ell)}(\Phi_t)\bigr\rangle_{\mathcal{H}_k^{\otimes \ell}}.
\end{equation}

\paragraph{Gram matrices.}
Let $\mathcal{T}^{\to}:=\{t\in\mathcal{T}: t+1\in\mathcal{T}\}$.
For $i,j\in\mathcal{T}^{\to}$, define
\begin{equation}\label{eq:gram_sig}
G_{ij}:=\kappa_{\mathrm{sig}}^{(n,\lambda)}(X_i,X_j),\qquad
A_{ij}:=\kappa_{\mathrm{sig}}^{(n,\lambda)}(X_i,X_{j+1}).
\end{equation}
Computational details are given in Appendix~\ref{app:signature}.

\subsection{Koopman matrix estimation by kEDMD}\label{sec:koopman_full}

Kernel EDMD yields a finite-dimensional approximation of the path-shift Koopman operator via the generalized eigenproblem
\begin{equation}\label{eq:gedmd_gen_eig}
A v = \mu\, G v,
\end{equation}
where $(\mu,v)$ approximate the Koopman eigenvalues and eigenfunctions in the RKHS induced by $\kappa_{\mathrm{sig}}$.

\paragraph{Orthogonalization and Koopman matrix.}
We stabilize \eqref{eq:gedmd_gen_eig} by orthogonalization based on the symmetric Gram matrix $G$.
Let $G=Q\Sigma^{2}Q^\top$ be an eigendecomposition and retain a numerical rank-$r$ subspace.
Define
\begin{equation}\label{eq:K_full}
K := \Sigma^{\dagger}Q^\top A Q \Sigma^{\dagger}\in\R^{r\times r}.
\end{equation}
This construction is algebraically equivalent to applying EDMD with an explicit feature map whose inner products
are given by the kernel \citep{williams2015}.

\paragraph{Sum-of-Pairs Kernel (SPK) baseline.}
As a snapshot-aligned baseline, we run the same kEDMD pipeline but replace $\kappa_{\mathrm{sig}}^{(n,\lambda)}$ with
\begin{equation}\label{eq:spk_kernel}
\kappa_{\mathrm{SPK}}(X_s,X_t)
:=\sum_{i=0}^{11} k\!\left(x'_{m_0+12s+i},\,x'_{m_0+12t+i}\right),
\end{equation}
where $k$ is the same base kernel.

\noindent\emph{SPK vs.\ signature kernel (interaction structure).}
Let $\{a_i\}_{i=1}^{12},\{b_p\}_{p=1}^{12}\subset\mathcal H$.
The SPK (direct-sum) inner product keeps only diagonal month-to-month terms:
\[
\Big\langle \bigoplus_{i=1}^{12} a_i,\ \bigoplus_{p=1}^{12} b_p \Big\rangle
=\sum_{i=1}^{12}\langle a_i,b_i\rangle .
\]
In contrast, the signature feature is the tensor-algebra exponential
$e^{a}:=\sum_{m\ge0} a^{\otimes m}/m!$ and
$\mathbf S(a_{1:12})=\bigotimes_{i=1}^{12} e^{a_i}$.
It expands by levels as
\[
\mathbf S(a_{1:12})
=
1
\oplus \Big(\sum_{i=1}^{12} a_i\Big)
\oplus
\Big(\frac12\sum_{i=1}^{12} a_i\otimes a_i+\sum_{1\le i<j\le 12} a_i\otimes a_j\Big)
\oplus 
\cdots ,
\]
and similarly for $\mathbf S(b_{1:12})$.
With the graded inner product on $T(\mathcal H)=\bigoplus_{m\ge0}\mathcal H^{\otimes m}$
(levels are orthogonal),
\begin{multline}
\big\langle \mathbf S(a_{1:12}),\mathbf S(b_{1:12})\big\rangle
=\langle 1,1\rangle
+\Big\langle \sum_{i=1}^{12} a_i,\ \sum_{p=1}^{12} b_p \Big\rangle \\
+\Big\langle
\frac12\sum_{i=1}^{12} a_i\otimes a_i+\sum_{1\le i<j\le 12} a_i\otimes a_j,\,
\frac12\sum_{p=1}^{12} b_p\otimes b_p+\sum_{1\le p<q\le 12} b_p\otimes b_q
\Big\rangle
+\cdots .
\end{multline}
Thus, beyond the level-one term, the inner product includes genuinely \emph{ordered} cross-month interactions through the $i<j$ contributions (and their higher-order analogues), whereas SPK excludes all such interactions by construction.

\paragraph{Climatology baseline.}
We also compare against an anchor-specific monthly climatology baseline; its anomaly-space definition is given in
Appendix~\ref{app:cv}.

\subsection{Evaluation protocols: LFO and LSO}\label{sec:protocols}

\paragraph{Notation.}
We use $t$ for the annual index of the segmented record and denote one-step transitions by $(X_t,X_{t+1})$.
For evaluation, we denote the forecast/validation anchor by $t_0$ and the lead time (in years) by $s$;
therefore, verification is performed at $t_0+s$.

\paragraph{Admissible anchors.}
Anchors are restricted to those for which the verification target exists and the required training data are available
under each protocol; the exact index sets follow the implementation (Appendix~\ref{app:lso_lfo}).

\paragraph{Two protocol roles.}
We use two strictly time-ordered protocols with distinct roles:
(i) \textbf{LFO} for forecasting cross-validation and out-of-sample skill evaluation;
(ii) \textbf{LSO} for hyperparameter selection for spectral analysis, followed by fitting a final matrix $K$ on all available transitions for reporting modes.

\paragraph{Leave-Future-Out (LFO): forecasting CV / skill.}
For each anchor $t_0$, we estimate a Koopman matrix $K^{[t_0]}$ using only one-step transitions that are strictly prior to the anchor
and then form an $s$-step prediction by iterating the learned one-year map starting from $X_{t_0}$.
We evaluate the prediction against the held-out target $X_{t_0+s}$.
The precise definition of the training transition set is given in Appendix~\ref{app:lso_lfo}.

\paragraph{Leave-$s$-Out (LSO): hyperparameter selection for modes.}
For each anchor $t_0$ and lead $s$, we re-estimate $K^{[t_0]}$ after excluding a contiguous forward block of one-step transitions around the anchor,
and aggregate the resulting diagnostics over anchors.
Hyperparameters are selected by aggregating the criterion described in Appendix~\ref{app:cv} over anchors.
After selection, we fit a \emph{final} Koopman matrix $K$ using all available one-step transitions in the record and report its spectrum and representative modes.
The precise definition of the excluded transitions is given in Appendix~\ref{app:lso_lfo}.

\paragraph{Baselines.}
We compare against (i) an anchor-specific monthly climatology forecast and (ii) the SPK-kEDMD baseline.
The climatology baseline is defined to use only information available at each forecast anchor and is evaluated in anomaly space;
see Appendix~\ref{app:cv}.
Details of the $s$-step forecast construction and the LSO splits are given in Appendix~\ref{app:sstep_lso}.

\section{Data and Experimental Design}\label{sec:exp}

\subsection{SST dataset}\label{sec:data}
We use the NOAA Extended Reconstructed Sea Surface Temperature dataset (ERSSTv5) \citep{ERSSTv5}.
The data are monthly SST fields on a $2^{\circ}\times 2^{\circ}$ grid covering January 1854--December 2022.
ERSSTv5 is a reconstructed product; prior to the satellite era, sparse observations required statistical reconstruction,
which may introduce dataset-specific artifacts. We treat ERSSTv5 as the reference target and present this study as a
proof-of-method. All methods are trained and evaluated on the same gridded product under the same time-ordered protocol.

\subsection{Annual segmentation and start-month choice}\label{sec:segmentation}
Annual path segments are constructed from monthly anomalies as described in Section~\ref{sec:preprocess_path}.
The segmentation depends on the start month $m_0$.
In the experiments below, we fix $m_0$ to August. The start month is selected by a preliminary LFO screening over the
12 calendar months.

\subsection{Protocol settings and hyperparameters}\label{sec:protocol_design}
All forecasting and spectral diagnostics use strictly time-ordered splits.
Forecast skill is evaluated with LFO and spectral diagnostics using LSO with a 5-year lead ($s=5$).
In the signature-kernel case, we tune the dilation $\lambda$ and the number of retained modes  (via the number of discarded conjugate pairs  $q$);
in the SPK baseline, only the number of modes is tuned. The cross-validation objective and anchor definitions are given in
Appendix~\ref{app:cv}. The signature truncation level is fixed to $n=7$.

\subsection{Skill metrics}\label{sec:metrics}
Forecast skill is evaluated in anomaly space.
Let $y,\hat y\in\R^{d}$ denote the reference and predicted anomaly fields at a verification time, defined on valid ocean
grid points. Let $w_i\ge 0$ be area weights with $\sum_i w_i = 1$ (in practice $w_i\propto \cos(\mathrm{lat}_i)$ and then
normalized over valid points).

\paragraph{Area-weighted RMSE.}
\begin{equation}\label{eq:rmse_def}
\mathrm{RMSE}(y,\hat y)
:=
\left(
\sum_{i} w_i\,(y_i-\hat y_i)^2
\right)^{1/2}.
\end{equation}

\paragraph{Kernel pattern correlation (kPC).}
We evaluate forecast skill in the RKHS induced by a positive semidefinite kernel $\kappa(\cdot,\cdot)$ on the prediction
targets (annual paths in our case). For a reference target $Y$ and a prediction $\widehat Y$, define
\begin{equation}\label{eq:kpc_kernel_def}
\mathrm{kPC}(Y,\widehat Y)
:=
\frac{\kappa(Y,\widehat Y)}
{\sqrt{\kappa(Y,Y)\,\kappa(\widehat Y,\widehat Y)}}.
\end{equation}
Thus, the cross-validation objective in \eqref{eq:cv_objective_app} is the negative mean of $\mathrm{kPC}$ over validation
anchors.

\section{Results}\label{sec:results}

This section reports forecasting performance under LFO and spectral diagnostics under LSO.
Unless stated otherwise, all evaluations compare the signature-kernel kEDMD pipeline (SigK-EDMD) against the climatology
baseline and the SPK-kEDMD baseline using identical preprocessing and data separation rules.
Forecast skill is quantified by area-weighted RMSE and kPC defined in Section~\ref{sec:metrics}.

\subsection{LFO forecast skill}\label{sec:results_lfo}

\begin{figure*}[t]
\centering
\includegraphics[width=0.49\textwidth]{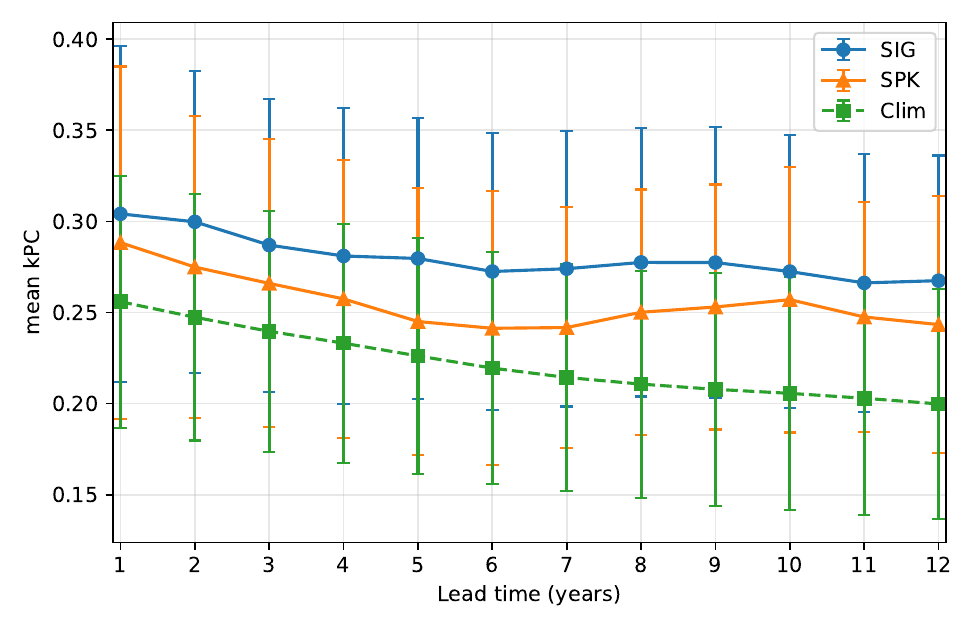}
\includegraphics[width=0.49\textwidth]{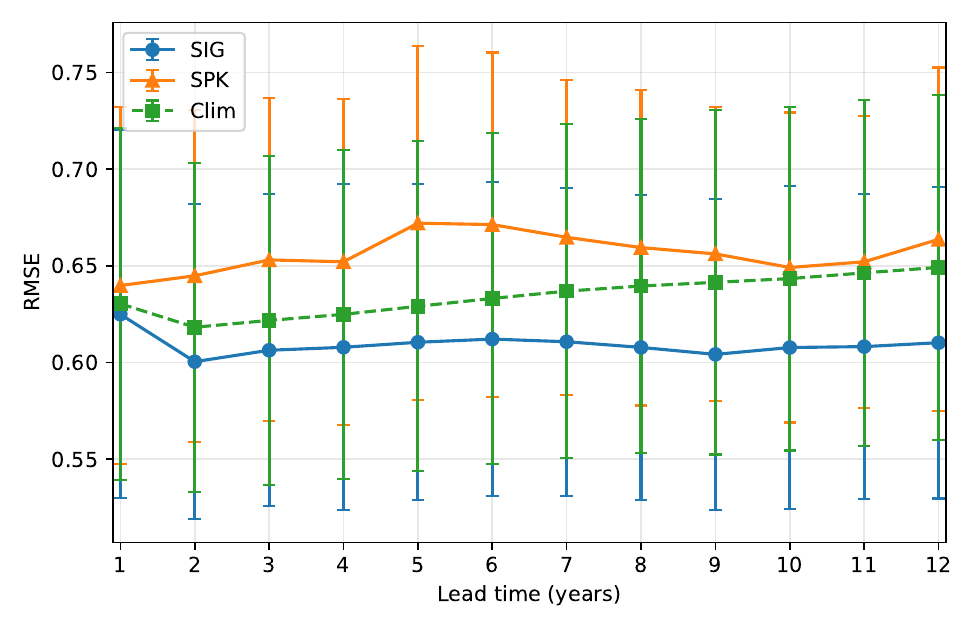}
\caption{Leave-future-out (LFO) forecast skill as a function of lead time (August-start).
Left: kPC. Right:  RMSE.}
\label{fig:lfo_leadtime_aug}
\end{figure*}

Figure~\ref{fig:lfo_leadtime_aug} summarizes the leave-future-out (LFO) forecast skill as a function of lead time for the August-start segmentation.
Across all lead times ($s=1$--12~yr), SigK-EDMD achieves the best performance in terms of both kPC (Eq.\,\eqref{eq:mean_kPC}) and RMSE (Eq.\,\eqref{eq:mean_RMS}).
The advantage is smallest at $s=1$~yr, whereas it becomes more pronounced at multi-year to decadal lead times.
Because the discrete-time Koopman analysis uses a 1~yr sampling interval, oscillatory modes in the spectrum with period shorter than 2~yr are not resolvable.

The SPK baseline improves kPC relative to climatology but often yields worse RMSE, indicating that month-wise kernel similarity does not necessarily translate into field-wise error reduction.
By contrast, the consistent gain of SigK-EDMD over SPK suggests that respecting within-year temporal order in the trajectory kernel provides a more informative lifting for the one-year shift operator.

\begin{figure*}[t]
\centering
\includegraphics[width=0.49\linewidth]{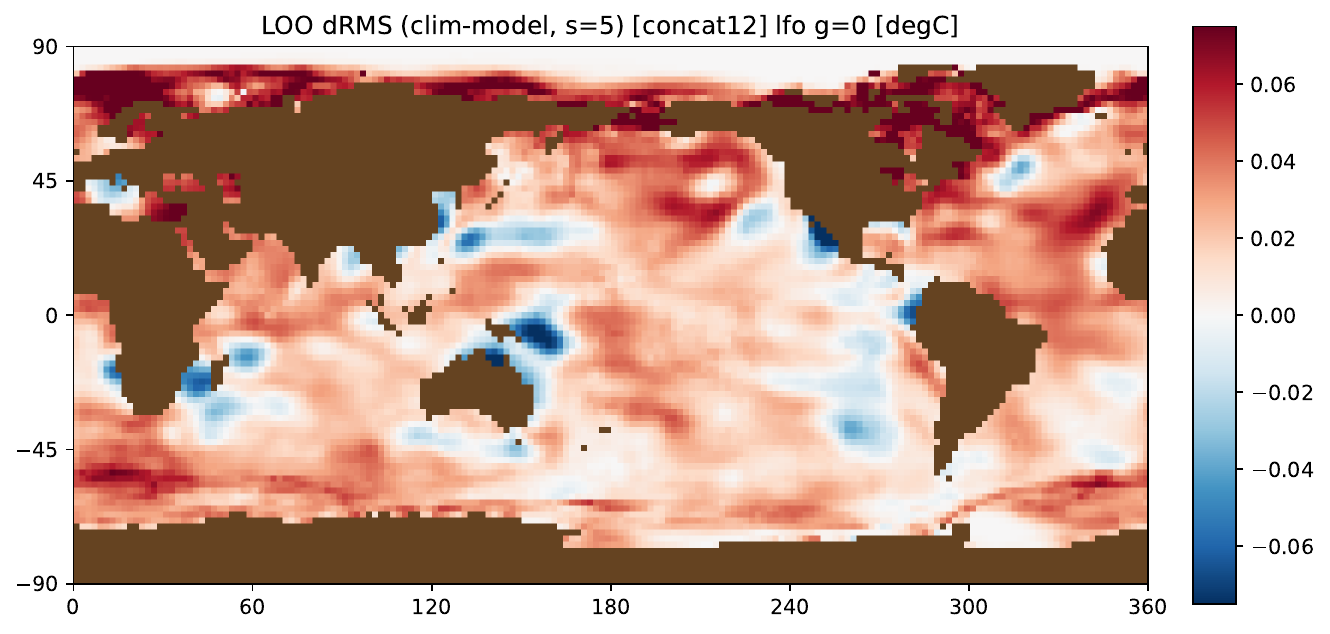}\hfill
\includegraphics[width=0.49\linewidth]{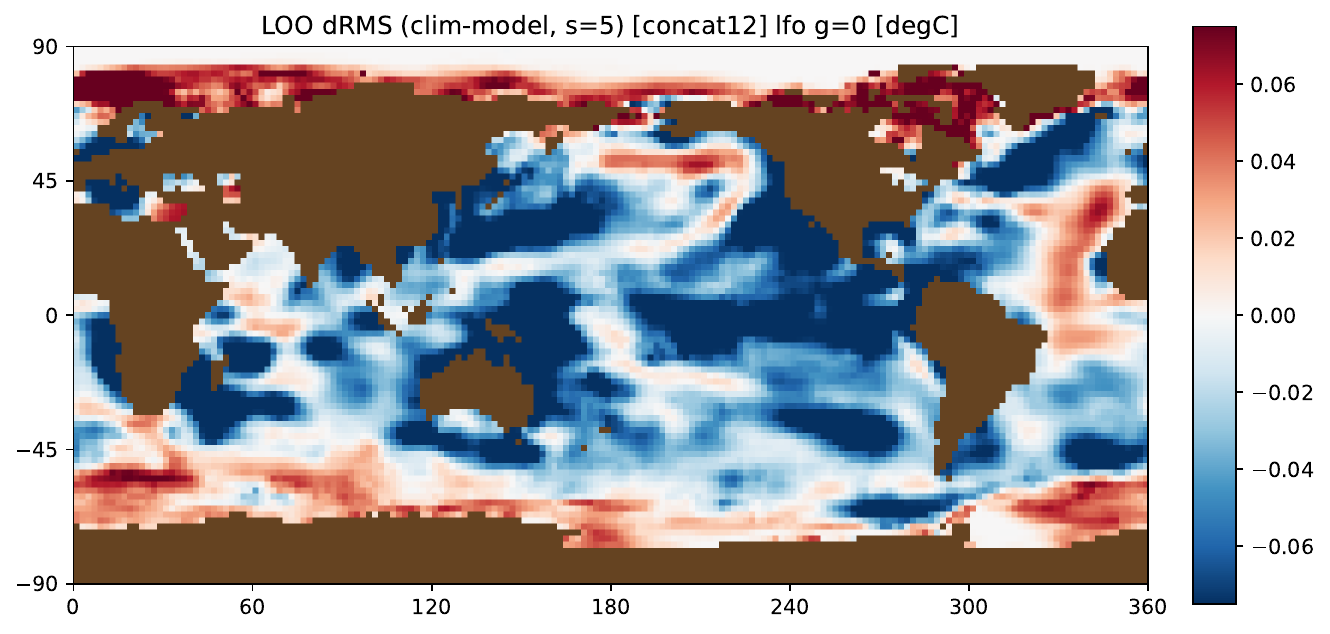}
\caption{Spatial distribution of RMSE difference relative to climatology at $s=5$ years (August-start, LFO).
We plot $\Delta\mathrm{RMSE}=\mathrm{RMSE}_{\text{clim}}-\mathrm{RMSE}_{\text{method}}$ (in $\si{\celsius}$);
positive values indicate improvement over climatology.
Left: SigK-EDMD. Right: SPK.}
\label{fig:drmse_maps}
\end{figure*}

Figure~\ref{fig:drmse_maps} visualizes the spatial distribution of skill at $s=5$~yr using
$\Delta\mathrm{RMSE}=\mathrm{RMSE}_{\mathrm{clim}}-\mathrm{RMSE}_{\mathrm{method}}$.
For SigK-EDMD, improvements are spatially heterogeneous and are more evident outside the tropical Pacific, including parts of the extratropical North Pacific, the Atlantic, and the Indian Ocean.
By contrast, SPK shows limited improvement and degrades RMSE over broad regions.

\subsection{LSO spectral diagnostics and representative modes}\label{sec:results_lso}
\begin{figure*}[t]
\centering
\includegraphics[width=0.49\textwidth]{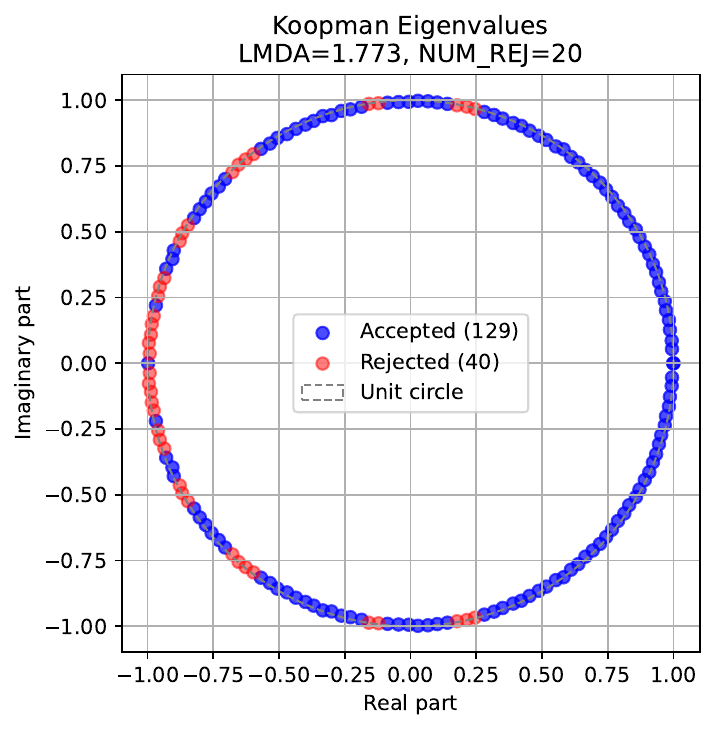}
\includegraphics[width=0.49\textwidth]{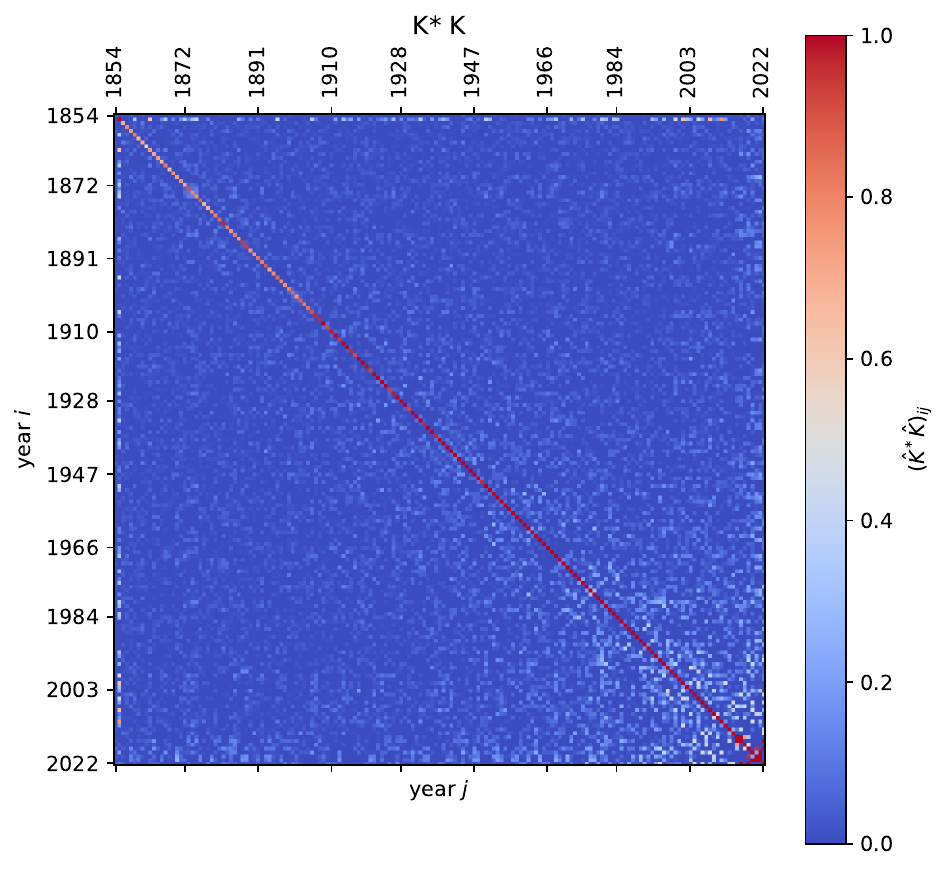}
\caption{Spectral diagnostics under LSO (August-start).
Left: Koopman eigenvalues of $K$ (unit circle as reference). Right: heatmap of $K^\ast K$.}
\label{fig:lso_spectral_aug}
\end{figure*}

Figure~\ref{fig:lso_spectral_aug} displays the Koopman eigenvalues of the estimated matrix $K$ (left) and $K^\ast K$ (right).
The eigenvalues are concentrated on or slightly inside the unit circle, indicating that the extracted oscillatory modes are
highly persistent (weakly damped) over interannual to decadal time scales.
Moreover, the matrix $K^\ast K$ is close to the identity, suggesting that although the dynamics are weakly dissipative,
the learned operator remains approximately unitary (i.e., nearly orthogonal) on the retained subspace.

To interpret each eigenvalue in terms of oscillation and decay, we convert the discrete-time eigenvalue $\mu_k$ into a
continuous-time rate via
\begin{equation}\label{eq:cont_rate_from_mu}
\frac{1}{\Delta t}\log(\mu_k)=\sigma_k+i\omega_k,
\qquad \Delta t=1~\mathrm{yr},
\end{equation}
where $\log(\cdot)$ denotes the complex logarithm (principal branch).
Then, the oscillation period and the e-folding decay time are
\begin{equation}\label{eq:period_efold}
T_k=\frac{2\pi}{\omega_k},
\qquad
\tau_k=\frac{1}{-\sigma_k}\quad(\sigma_k<0).
\end{equation}

\begin{figure*}[t]
\centering
\includegraphics[width=0.64\linewidth]{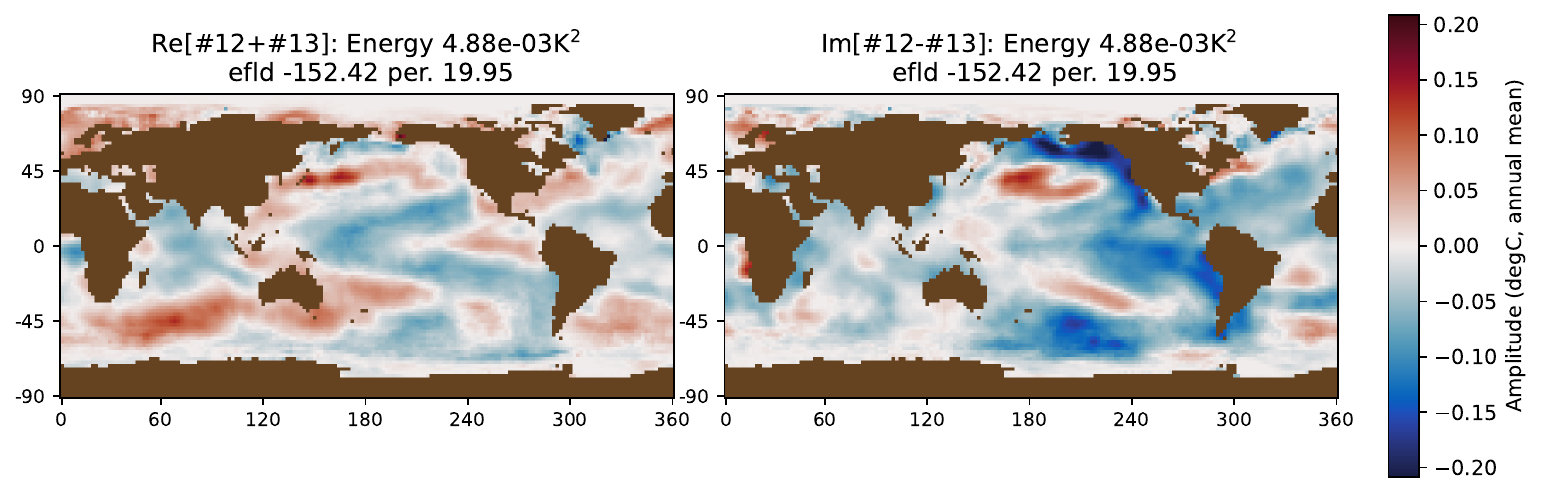}\hfill
\includegraphics[width=0.34\linewidth]{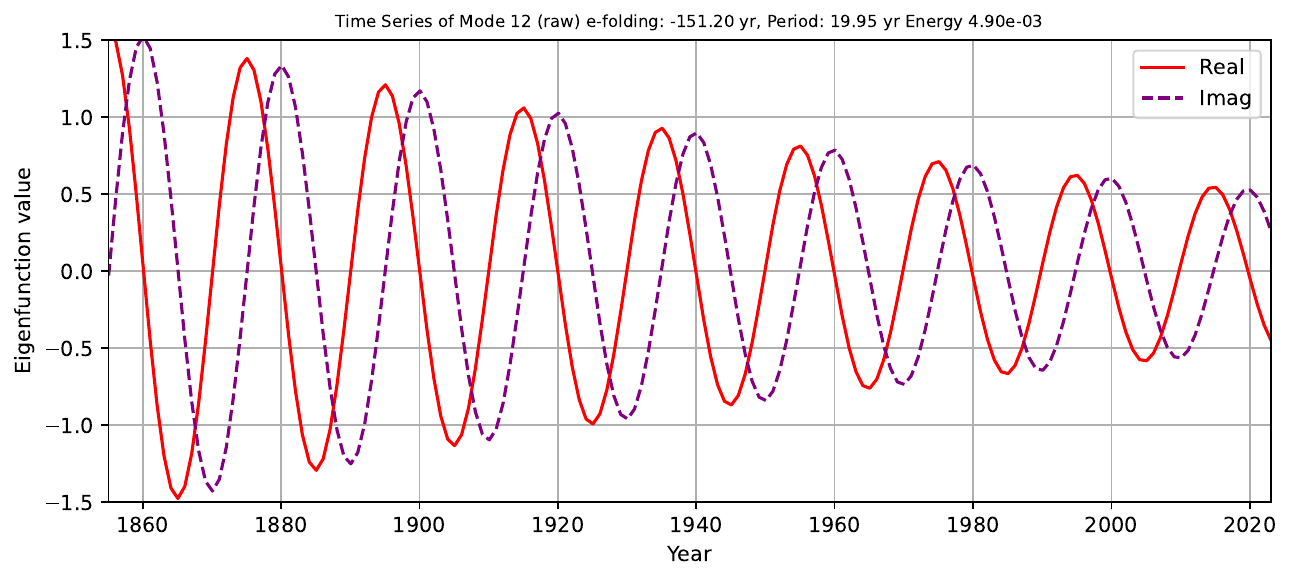}\\[0.6ex]
\includegraphics[width=0.64\linewidth]{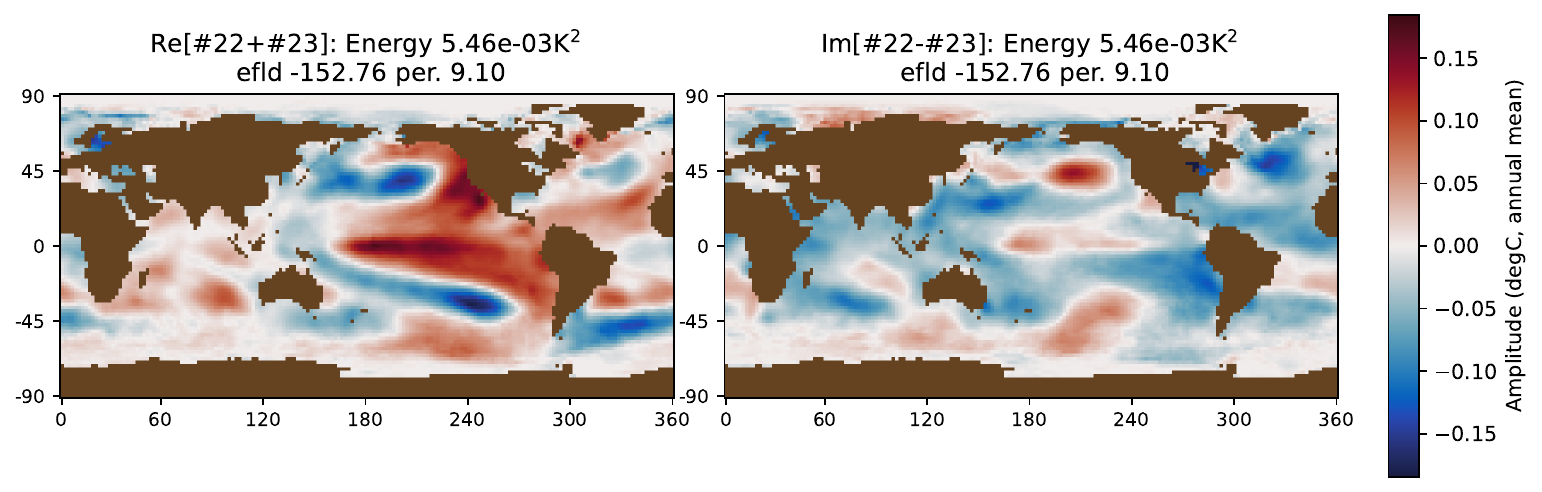}\hfill
\includegraphics[width=0.34\linewidth]{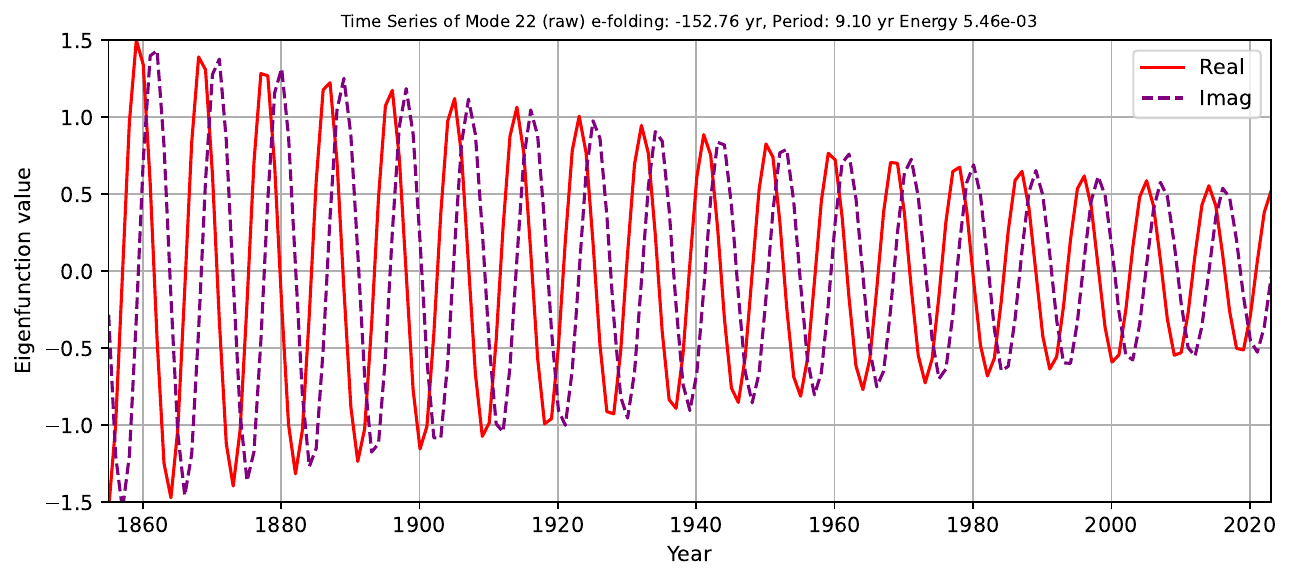}\\[0.6ex]
\includegraphics[width=0.64\linewidth]{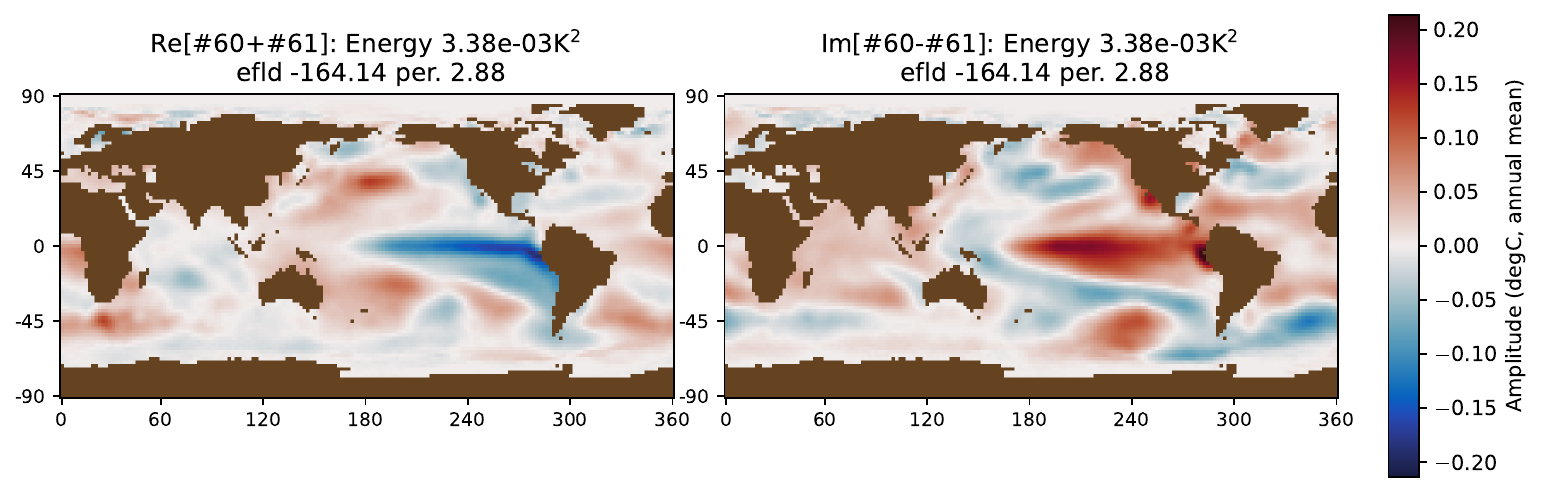}\hfill
\includegraphics[width=0.34\linewidth]{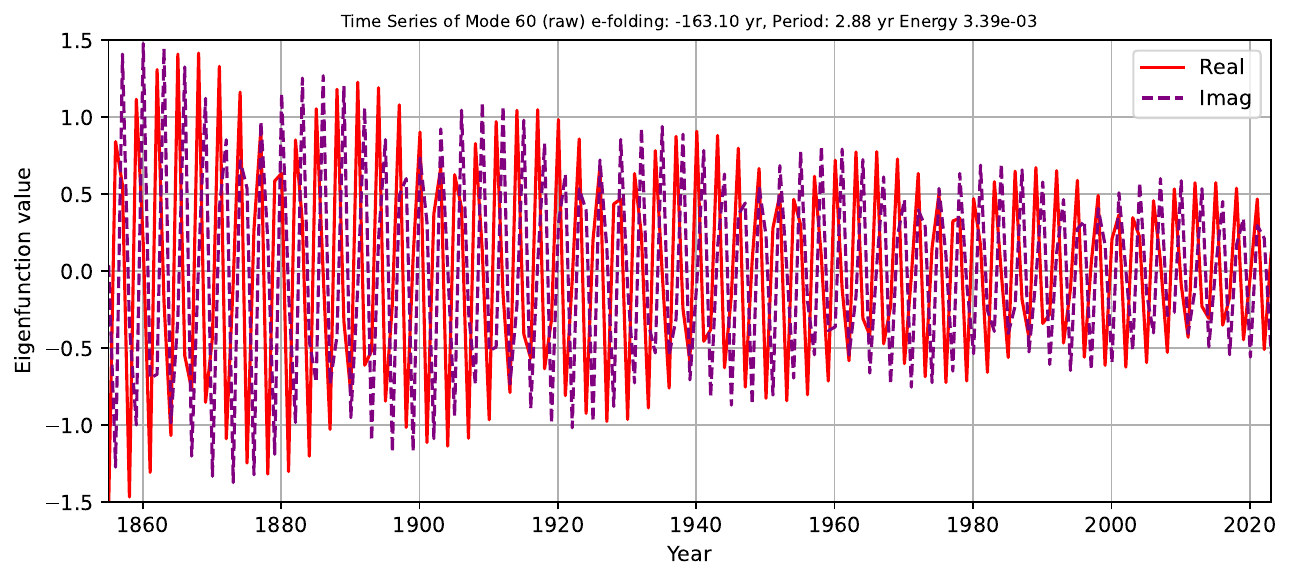}
\caption{Selected Koopman modes under LSO (August-start, $s=5$):
mode~\#12 (period $\approx 20$~yr; KOE-like), \#22 (period $\approx 9.1$~yr; PDO-like), and \#60 (period $\approx 2.9$~yr; CP-ENSO-like).
These labels indicate qualitative resemblance to commonly discussed SST-variability patterns.
Right: corresponding mode coordinate along the record, shown in a de-amplified form for readability.
The plotted Koopman time series shows eigenfunction values $a_k(t)=\psi_k(X_t)$ with eigenvalue $\mu_k$; any decay when $|\mu_k|<1$ reflects the fitted finite-dimensional approximation (finite samples and truncation), not physical dissipation.}
\label{fig:koopman_modes_three}
\end{figure*}

Using \eqref{eq:cont_rate_from_mu}--\eqref{eq:period_efold}, we summarize the dominant oscillatory eigenmodes by their
period and decay characteristics, and visualize their associated spatial patterns.
Figure~\ref{fig:koopman_modes_three} shows representative oscillatory Koopman modes.
\paragraph{Mode visualization space.}
In LSO, the Koopman modes are estimated from annual paths; however, the spatial maps in Fig.~\ref{fig:koopman_modes_three}
are shown after projection onto the \emph{annual-mean} SST anomaly field for each year, to focus on long-term (year-to-year) variability.
Concretely, by letting
\[
\bar x'_t := \frac{1}{12}\sum_{i=0}^{11} x'_{m_0+12t+i}\in\R^{d},
\]
we visualize each mode in the $\bar x'$-space by applying the same annual-mean projection to the reconstructed anomaly
fields and plot the real and imaginary parts.

\paragraph{Representative modes.}
Among the oscillatory modes (Appendix~\ref{app:spectrum}) under the August-start LSO setting ($s=5$),
we highlight three modes whose spatial patterns are qualitatively similar to commonly discussed SST-variability structures:
mode~\#12 (period $\approx 20$~yr) resembles a Kuroshio--Oyashio Extension (KOE)-like pattern in \citet{DiLorenzo2023PDV};
mode~\#22 (period $\approx 9.1$~yr) resembles a Pacific Decadal Oscillation (PDO)-like pattern in \citet{DiLorenzo2023PDV}
or the $\sim 9.22$-yr mode in \citet{Navarra2024};
and mode~\#60 (period $\approx 2.9$~yr) resembles a central-Pacific (CP) ENSO-like pattern in \citet{DiLorenzo2023PDV}
or the $\sim 2.86$-yr mode in \citet{Navarra2024}.
These labels are used only as descriptive shorthand based on qualitative pattern similarity; we adopt the terms ``KOE,'' ``PDO,'' and ``CP-ENSO'' from \citet{DiLorenzo2023PDV}.
A detailed physical interpretation of the extracted modes is beyond the scope of this methodological study and is left for future work.

Overall, beyond these three examples, the extracted oscillatory modes span multiple time scales and include patterns qualitatively similar to previously reported SST-variability structures (e.g., \citet{DiLorenzo2023PDV,Navarra2024}).
Although we do not attempt a detailed physical interpretation, this cross-study pattern consistency supports the plausibility of the estimated spectrum and the associated spectral diagnostics.

\section{Conclusion}\label{sec:conc}
We applied a \emph{path-based} Koopman approach to a high-dimensional geophysical time series, using SST as a concrete example. To construct a data-driven function space of observables on annual trajectories, we computed signature-kernel Gram matrices for pairs of annual SST paths and established a fully kernelized kEDMD procedure to obtain a finite-dimensional Koopman matrix. Accordingly, we made the following observations:
\begin{itemize}
\item Under strictly time-ordered cross-validation, the proposed method achieved higher forecast skill than both a month-wise snapshot-matching kernel baseline and an anchor-based climatology baseline.
\item From the estimated Koopman matrix, we extracted intrinsic oscillatory modes of the SST record as tuples of (period, decay rate, and complex spatial mode structure), enabling unified spectral diagnostics within the same estimator.
\item The proposed framework provides a nonlinear data-analysis tool for identifying climate-relevant multi-year variability from SST alone, and it is directly applicable to other high-dimensional time series where history dependence is important.
\end{itemize}
We note that ERSSTv5 is a statistically reconstructed product; especially prior to the satellite era, sparse observations necessitate statistical reconstruction, and the gridded fields should not be conflated with direct measurements.
Accordingly, the extracted modes should be interpreted as modes of the reconstructed product rather than as direct observational facts.

\noindent\textbf{Code availability.}
The code used in this study is archived on Zenodo (v1.0.3, DOI: \url{https://doi.org/10.5281/zenodo.18736428}).
The development repository is available at \url{https://github.com/nozomi-sugiura/ChronoEDMD}.

\medskip
\noindent\textbf{Data availability.}
The sea surface temperature dataset analyzed in this study is NOAA ERSSTv5, available from the NOAA Physical Sciences Laboratory at
\url{https://psl.noaa.gov/data/gridded/data.noaa.ersst.v5.html}.

\begin{acknowledgements}

\end{acknowledgements}

\section*{Declarations}
\noindent\textbf{Conflict of interest.} The author declares that there is no conflict of interest.

\medskip
\noindent\textbf{Ethical approval.} Not applicable.

\bibliographystyle{spbasic}
\bibliography{koopman}

\appendix

\section{Preprocessing: rolling climatology, anomalies, and annual paths}\label{app:preprocess}

\paragraph{Monthly SST and rolling climatology.}
Let $\{x_m\}_{m=m_{\min}}^{m_{\max}}$ be monthly SST fields reshaped into vectors $x_m\in\R^{d}$.
Fix a start-month index $m_0$ and define the month-of-year label
\[
r(m):=(m-m_0)\ (\bmod~12)\in\{0,1,\dots,11\}.
\]
For each month $m$, define a strictly past-only rolling climatology by averaging available past
values of the same calendar month with a 12-month stride and a window of at most 30 years:
\begin{equation}\label{eq:app_climatology_rolling}
c_m := \frac{1}{N(m)}\sum_{j=0}^{N(m)-1} x_{m-12j},\qquad
N(m):=\min\!\left(30,\; \left\lfloor\frac{m-m_{\min}}{12}\right\rfloor+1\right).
\end{equation}

\paragraph{Anomalies.}
\begin{equation}\label{eq:app_anomaly}
x'_m := x_m - c_m \in\R^{d}.
\end{equation}

\paragraph{Annual paths via cumulative sums.}
For year index $t$, define cumulative nodes
\begin{equation}\label{eq:app_cumsum_nodes}
X_{t,0}:=0,\qquad
X_{t,k}:=\sum_{i=0}^{k-1} x'_{m_0+12t+i}\quad (k=1,\dots,12),
\end{equation}
and let $X_t:[0,1]\to\R^{d}$ be the piecewise-linear interpolation through $\{X_{t,k}\}_{k=0}^{12}$.

\paragraph{One-year shift on available pairs.}
Let $\mathcal{T}$ be the set of years for which segments exist and $\mathcal{T}^{\to}:=\{t\in\mathcal{T}:t+1\in\mathcal{T}\}$.
Define $F(X_t):=X_{t+1}$ for $t\in\mathcal{T}^{\to}$.

\section{Signature-kernel recursion induced by a base kernel}\label{app:signature}

This appendix records the recursion used to compute the truncated signature kernel induced by a base kernel,
following \citet{Kiraly2019}. Only the required notation and recursion are given.

\subsection{Base kernel}\label{app:base_kernel}

\paragraph{Area-weighted RBF kernel.}
Let $w_i\ge 0$ be area weights over valid ocean grid points with $\sum_i w_i = 1$.
On a regular latitude--longitude grid, we take preliminary weights proportional to $\cos(\mathrm{lat}_i)$ and then normalize.
Define the weighted squared distance
\begin{equation}\label{eq:weighted_norm}
\|x-y\|_{w}^{2}:=\sum_{i} w_i (x_i-y_i)^2 .
\end{equation}
We use an area-weighted radial basis function (RBF) kernel
\begin{equation}\label{eq:base_k}
k(x,y)=C\,\exp\!\left(-\frac{\|x-y\|_{w}^{2}}{2\sigma^2}\right),
\qquad C>0,
\end{equation}
where we set $C=1$ without loss of generality, since a multiplicative constant
in the base kernel can be absorbed into the dilation parameter in the truncated
signature-kernel expansion (Eq.~\eqref{eq:sig_kernel}), i.e., $\lambda\mapsto\lambda\sqrt{C}$.

\paragraph{Choice of $\sigma$.}
The scale $\sigma$ is fixed once from the full record using the same weighted geometry \eqref{eq:weighted_norm}
and then held constant throughout all experiments.
In the signature-kernel case, $\sigma$ is computed from the set of path nodes $\{X_{t,k}\}$.
For the SPK baseline, $\sigma$ is computed from monthly anomaly snapshots $\{x'_{m_0+12t+i}\}$.

\subsection{Kernel recursion}
Let $x,y$ be piecewise-linear paths with nodes $\{X_i\}_{i=0}^{N}$ and $\{Y_j\}_{j=0}^{M}$.
Define
\[
\Delta k_{i,j}:=k(X_{i+1},Y_{j+1})-k(X_{i+1},Y_{j})-k(X_{i},Y_{j+1})+k(X_{i},Y_{j}).
\]
Let $\kappa^{i,j}_{0}(x,y)=1$. The discrete recursion for the truncated signature kernel is given by
\begin{equation}\label{eq:sigk_rec_disc}
\kappa^{i,j}_{m+1}(x,y)
= 1+\sum_{p=0}^{i-1}\sum_{q=0}^{j-1}\kappa^{p,q}_{m}(x,y)\,\Delta k_{p,q},
\qquad m\ge 0.
\end{equation}

\section{Cross-validation objective and baseline definitions}\label{app:cv}

\subsection{Hyperparameters}
In the signature-kernel case, we tune the number of discarded conjugate pairs  $q$ (hence the retained numerical rank) and dilation $\lambda$.
In the SPK baseline, $\lambda$ is not used, and we tune only $q$.

\subsection{Kernel-alignment objective}
Let $\mathcal{A}_{\mathrm{val}}$ denote the set of validation anchors under a time-ordered protocol.
For each anchor $t_0\in\mathcal{A}_{\mathrm{val}}$, let $X^{(q,\lambda)}_{\mathrm{pred},t_0}$ denote the $s$-step predicted path
and $X_{\mathrm{true},t_0}$ the corresponding reference path.
We select hyperparameters by minimizing
\begin{equation}\label{eq:cv_objective_app}
J(q,\lambda)
:=
-\frac{1}{|\mathcal{A}_{\mathrm{val}}|}
\sum_{t_0\in\mathcal{A}_{\mathrm{val}}}
\frac{\kappa^{(\lambda)}\!\left(X^{(q,\lambda)}_{\mathrm{pred},t_0},\,X_{\mathrm{true},t_0}\right)}
{\sqrt{\kappa^{(\lambda)}\!\left(X^{(q,\lambda)}_{\mathrm{pred},t_0},\,X^{(q,\lambda)}_{\mathrm{pred},t_0}\right)\,
       \kappa^{(\lambda)}\!\left(X_{\mathrm{true},t_0},\,X_{\mathrm{true},t_0}\right)}}.
\end{equation}

\subsection{Climatology baseline in anomaly space}
At anchor year $t_0$ and lead $s$, the climatology forecast is defined to use only information available at the anchor:
it repeats the anchor-year month-of-year climatology,
\[
\widehat{x}^{\mathrm{clim}}_{m_0+12(t_0+s)+i}:=c_{m_0+12t_0+i},\qquad i=0,\dots,11.
\]
Evaluation is performed in anomaly space, where $x'_m=x_m-c_m$ is defined using the (rolling) climatology $c_m$
at the verification month $m$. Therefore, the corresponding baseline anomalies are
\begin{equation}\label{eq:clim_baseline_anom_app}
\widehat{x}'^{\,\mathrm{clim}}_{m_0+12(t_0+s)+i}
= c_{m_0+12t_0+i}-c_{m_0+12(t_0+s)+i},
\qquad i=0,\dots,11,
\end{equation}
i.e., the climatology baseline expressed in the verification-month anomaly coordinates.

\section{\texorpdfstring{$s$}{s}-step forecasting and Leave-\texorpdfstring{$s$}{s}-out evaluation}
\label{app:sstep_lso}

\paragraph{Notation.}
The split index $t_0$ in this appendix corresponds to the anchor $t_0$ used in the main text.

This appendix specifies (i) $s$-step (i.e., $s$-year-ahead) forecasting based on the kEDMD Koopman matrix,
and (ii) the leave-$s$-out (LSO) splits used for spectral diagnostics.
Throughout, one ``step'' corresponds to \textbf{one year}; $s$ denotes the \textbf{lead time of $s$ years}.

\subsection{Objects for forecasting and verification}
\label{app:objects_forecast}

We work with annual \emph{paths} $\{X_t\}_{t=0}^{M-1}\subset C([0,1],\R^{d})$ defined from monthly anomalies
(Appendix~\ref{app:preprocess}). For completeness, we denote the within-year anomaly snapshots by
$\{x'_{m_0+12t+i}\}_{i=0}^{11}$ and the cumulative nodes by $\{X_{t,k}\}_{k=0}^{12}$.

\paragraph{Annual-state vector (used only for RMSE).}
To compute RMSE on anomaly fields, it is convenient to stack the 12 monthly anomaly snapshots into a single vector:
\begin{equation}\label{eq:app_state_stack}
z_t := \mathrm{vec}\!\left(x'_{m_0+12t+0},\,x'_{m_0+12t+1},\,\dots,\,x'_{m_0+12t+11}\right)\in\R^{D},
\qquad D:=12\,d.
\end{equation}

\subsection{Leave-\texorpdfstring{$s$}{s}-out splits and LFO start times}
\label{app:lso_lfo}

\paragraph{One-step transitions.}
We index one-step transitions by $t$ as $(X_t,X_{t+1})$.

\paragraph{Admissible anchors.}
We consider anchors $t_0$ such that the verification target $X_{t_0+s}$ exists.
Additionally, in the implementation, we skip the first $s$ years to avoid the initial preprocessing transient,
so the anchor set is defined as
\[
\mathcal{T}_{\mathrm{anchor}} := \{\, t_0 \in \{0,\dots,M-1\}\mid s \le t_0 \le M-s-1 \,\},
\]
where $M$ is the number of annual segments.

\paragraph{LFO training transitions (forecasting CV / skill).}
For a given anchor $t_0\in\mathcal{T}_{\mathrm{anchor}}$, LFO uses only transitions strictly preceding the anchor for training:
\[
\mathcal{I}^{[t_0]\to}_{\mathrm{LFO}}
:= \{\, t \in \{0,\dots,M-2\}\mid t \le t_0-1 \,\}.
\]
Equivalently, all transitions with $t\ge t_0$ are excluded.

\paragraph{LSO training transitions (spectral hyperparameter selection).}
For a given anchor $t_0\in\mathcal{T}_{\mathrm{anchor}}$ and lead $s$, LSO removes the contiguous forward block of transition indices
\[
\{t_0, t_0+1, \dots, t_0+s\},
\]
when these transitions exist, and trains on the remaining transitions:
\[
\mathcal{I}^{[t_0]\to}_{\mathrm{LSO}}
:= \{0,\dots,M-2\}\setminus\{t_0,\dots,t_0+s\}.
\]
Thus, LSO excludes the one-step transitions
\[
(X_{t_0},X_{t_0+1}), (X_{t_0+1},X_{t_0+2}), \dots, (X_{t_0+s},X_{t_0+s+1}),
\]
and uses all other available one-step transitions for estimating $K^{[t_0]}$.

\paragraph{Remark (consistency with Gram-matrix indexing).}
All Gram and cross-Gram matrices in Appendix~\ref{app:sstep_lso} are constructed from the annual paths
$\{X_t\}$ with transition indices $t$ restricted to $\mathcal{I}^{[t_0]\to}_{\mathrm{LFO}}$ (for LFO) or
$\mathcal{I}^{[t_0]\to}_{\mathrm{LSO}}$ (for LSO), respectively.

\subsection{kEDMD Koopman matrix for each split}
\label{app:kedmd_split}

Fix $t_0$ and let $\kappa(\cdot,\cdot)$ denote the training kernel on annual paths.
Define
\begin{equation}\label{eq:app_GA_split}
G^{[t_0]}_{tt'} := \kappa(X_t,X_{t'}) ,
\qquad
A^{[t_0]}_{tt'} := \kappa(X_t,X_{t'+1}),
\qquad (t,t'\in\mathcal{I}^{[t_0]\to}).
\end{equation}
Let a truncated eigendecomposition be
\begin{equation}\label{eq:app_G_eig}
G^{[t_0]} = Q^{[t_0]}\,\Sigma^{[t_0]2}\,Q^{[t_0]\top}.
\end{equation}
Then
\begin{equation}\label{eq:app_K_split}
K^{[t_0]}
= \Sigma^{[t_0]\dagger} Q^{[t_0]\top} A^{[t_0]} Q^{[t_0]} \Sigma^{[t_0]\dagger}.
\end{equation}

\subsection{\texorpdfstring{$s$}{s}-step forecasting}
\label{app:sstep_forecast}

Let the eigendecomposition of $K^{[t_0]}$ be
\begin{equation}\label{eq:app_eig}
K^{[t_0]}v^{[t_0]}_k = \mu^{[t_0]}_k\,v^{[t_0]}_k,
\qquad k=1,\dots,r,
\end{equation}
and set $V^{[t_0]} := (v^{[t_0]}_1,\dots,v^{[t_0]}_r)$.

\paragraph{Koopman modes in annual-state space.}
Let $Z^{[t_0]}$ be the data matrix whose rows are the stacked states $z_t^\top$ for $t\in\mathcal{I}^{[t_0]\to}$.
Define
\begin{align}
  \Phi_z^{[t_0]} &:= Q^{[t_0]}\Sigma^{[t_0]}V^{[t_0]},\\
  \Xi^{[t_0]} &:= (\Phi_z^{[t_0]})^{\dagger} Z^{[t_0]},
\end{align}
and denote by $\xi_k^{[t_0]}\in\R^{D}$ the $k$-th column of $\Xi^{[t_0]}$.

\paragraph{Forecast coefficients from the initial path.}
Let $X_{t_0}$ be the initial annual path (excluded from training under LSO).
Define the surrogate eigenfunction value at $X_{t_0}$ by
\begin{equation}\label{eq:app_psi}
\psi_k^{[t_0]}(X_{t_0})
:=
\sum_{t\in\mathcal{I}^{[t_0]\to}}
\kappa(X_{t_0},X_t)\,
\Bigl[\,Q^{[t_0]}\Sigma^{[t_0]\dagger}v^{[t_0]}_k\,\Bigr]_{t}.
\end{equation}

\paragraph{$s$-step prediction of the stacked annual anomaly state.}
The $s$-year-ahead prediction of the stacked annual anomaly state vector is
\begin{equation}\label{eq:app_zpred}
z_{t_0+s}^{\mathrm{pred}}
= \sum_{k=1}^{r}
  \bigl(\mu_k^{[t_0]}\bigr)^{s}\,
  \psi_k^{[t_0]}(X_{t_0})\,
  \xi_k^{[t_0]}
\in \R^{D},
\end{equation}
which is then split into 12 month blocks to obtain the predicted anomaly fields.

\paragraph{Month-wise anomaly-field prediction.}
Let $\xi^{[t_0]}_{k,i}\in\R^{d}$ be the restriction of $\xi^{[t_0]}_k\in\R^{d}$ to month block $i$ ($i=0,\dots,11$).
Then, the predicted monthly anomaly field at month index $m=m_0+12(t_0+s)+i$ is
\begin{equation}\label{eq:app_xprime_pred}
(x'_{m_0+12(t_0+s)+i})^{\mathrm{pred}}
= \sum_{k=1}^{r}
  \bigl(\mu_k^{[t_0]}\bigr)^{s}\,
  \psi_k^{[t_0]}(X_{t_0})\,
  \xi^{[t_0]}_{k,i},
\qquad i=0,1,\dots,11.
\end{equation}

\paragraph{Predicted path for kernel-based verification.}
From the predicted monthly anomalies, we define the predicted cumulative nodes
\begin{equation}\label{eq:app_path_pred}
X^{\mathrm{pred}}_{t_0+s,0} := 0,\qquad
X^{\mathrm{pred}}_{t_0+s,k} := \sum_{i=0}^{k-1} (x'_{m_0+12(t_0+s)+i})^{\mathrm{pred}}
\quad (k=1,\dots,12),
\end{equation}
and let $X^{\mathrm{pred}}_{t_0+s}$ be the corresponding piecewise-linear path.
Similarly, $X^{\mathrm{true}}_{t_0+s}$ is defined from the true anomalies.

\subsection{Mode selection via the resEDMD residual indicator}
\label{app:resdmd_mode_selection}

For each LSO split indexed by $t_0$, we compute a residual indicator for each eigenvalue of $K^{[t_0]}$
and discard the least reliable modes.

\paragraph{``Next--next'' matrix and whitening.}
We define the one-step ``next--next'' Gram matrix on the same training index set:
\begin{equation}\label{eq:app_L_nn}
L^{[t_0]}_{tt'} := \kappa(X_{t+1},X_{t'+1}),
\qquad (t,t'\in\mathcal{I}^{[t_0]\to}),
\end{equation}
and recall \eqref{eq:app_G_eig}.
The whitened ``next--next'' matrix is
\begin{equation}\label{eq:app_K2}
K^{[t_0]}_{2}
:= \Sigma^{[t_0]\dagger} Q^{[t_0]\top}\,L^{[t_0]}\,Q^{[t_0]}\Sigma^{[t_0]\dagger}.
\end{equation}

\paragraph{Residual indicator.}
Let $u^{[t_0]}_j$ be a left eigenvector of $K^{[t_0]}$, i.e.,
\begin{equation}\label{eq:app_left_eig}
K^{[t_0]\top} u^{[t_0]}_j = \mu^{[t_0]}_j\, u^{[t_0]}_j .
\end{equation}
Following \citet{colbrook2024}, define
\begin{equation}\label{eq:app_res_indicator}
\mathrm{res}^{[t_0]}_j
:=
\sqrt{
\max\!\left(
0,\,
\frac{\Re\!\bigl((u^{[t_0]}_j)^{*}\,K^{[t_0]}_{2}\,u^{[t_0]}_j\bigr)}
     {\Re\!\bigl((u^{[t_0]}_j)^{*}u^{[t_0]}_j\bigr)}
- \bigl|\mu^{[t_0]}_j\bigr|^2
\right)} .
\end{equation}

\paragraph{Filtering rule (discarding $q$ conjugate pairs).}
We group eigenvalues into complex-conjugate pairs, and each pair is assigned a score equal to the larger residual within the pair.
The $q$ pairs with the largest scores are then discarded, and the remaining modes are retained.

\subsection{Evaluation metrics for LSO}
\label{app:metrics_lso}

\paragraph{Area-weighted RMS on anomaly fields.}
Let $e^{[t_0]} := z_{t_0+s} - z_{t_0+s}^{\mathrm{pred}}\in\R^{D}$.
Let $w_\ell\ge 0$ denote spatial weights with $\sum_{\ell=1}^{d}w_\ell=1$ and extend them to the stacked state by repeating
the same spatial weights for each monthly block:
\[
W := I_{12}\otimes \mathrm{diag}(w_1,\dots,w_{d}).
\]
Define
\begin{equation}\label{eq:app_rms_aw}
\mathrm{RMS}^{[t_0]}
:= \left( (e^{[t_0]})^\top W\, e^{[t_0]} \right)^{1/2}.
\end{equation}

\paragraph{kPC on paths via the signature kernel.}
Let $\kappa_{\mathrm{sig}}$ be the signature kernel used for verification on annual paths.
Define
\begin{equation}\label{eq:app_kpc}
\mathrm{kPC}^{[t_0]}
:=
\frac{
\kappa_{\mathrm{sig}}\!\left(X^{\mathrm{true}}_{t_0+s},\,X^{\mathrm{pred}}_{t_0+s}\right)
}{
\sqrt{\kappa_{\mathrm{sig}}\!\left(X^{\mathrm{true}}_{t_0+s},\,X^{\mathrm{true}}_{t_0+s}\right)}
\sqrt{\kappa_{\mathrm{sig}}\!\left(X^{\mathrm{pred}}_{t_0+s},\,X^{\mathrm{pred}}_{t_0+s}\right)}
}.
\end{equation}

\subsection{Aggregation over validation starts}
\label{app:aggregation}
Let $\mathcal{T}$ be the set of admissible validation start years. We report the root-mean-squared error across anchors:
\begin{equation}\label{eq:mean_RMS}
\overline{\mathrm{RMS}}(s)
:= \left(\frac{1}{|\mathcal{T}|}\sum_{t_0\in\mathcal{T}} \bigl(\mathrm{RMS}^{[t_0]}\bigr)^2 \right)^{1/2}.
\end{equation}
We aggregate kPC by the arithmetic mean:
\begin{equation}
\overline{\mathrm{kPC}}(s)
:= \frac{1}{|\mathcal{T}|}\sum_{t_0\in\mathcal{T}} \mathrm{kPC}^{[t_0]}.
\label{eq:mean_kPC}
\end{equation}
\newpage

\section{Overview of the spectrum}\label{app:spectrum}
\begin{figure}[h]
\centering
\includegraphics[width=0.93\textwidth]{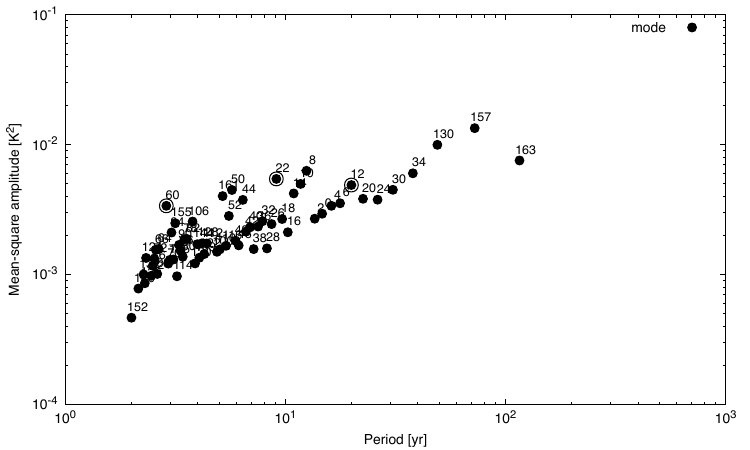}
\caption{Koopman spectrum of the learned one-year operator, shown as period (horizontal) versus area-weighted mean-square amplitude (vertical) on log--log axes.
Each point corresponds to a mode and is labeled by its mode index; circled points indicate the modes discussed in the main text (\#12, \#22, \#60).}
\label{fig:spectrum}
\end{figure}

Figure~\ref{fig:spectrum} summarizes the oscillatory Koopman modes of the learned one-year operator
as a period--amplitude scatter plot (log--log axes), where the vertical axis is the area-weighted mean-square amplitude [$\mathrm{K}^2$].
Because the operator is estimated on annual transitions ($\Delta t=1$~yr), periods shorter than $2$~yr are not resolvable (Nyquist limit).
In the main text we discuss three representative modes (\#12, \#22, and \#60), indicated in the figure.
Although we do not interpret the full spectrum here, the distribution suggests several apparent groupings (band-like structures) in period--amplitude space.
\end{document}